\documentclass[final,3p,times,sort&compress,twocolumn]{elsarticle}
\pdfoutput=1

\usepackage{graphicx}

\usepackage[usenames,dvipsnames]{xcolor}  
\usepackage{amssymb,amsfonts,amsmath}

\begin{document} 

\title {Ising-PageRank model of opinion formation on social networks}

\author[LPT]{Klaus M. Frahm}

\author[LPT]{Dima L.~Shepelyansky}

\address[LPT]{\mbox{Laboratoire de Physique Th\'eorique, IRSAMC, 
Universit\'e de Toulouse, CNRS, UPS, 31062 Toulouse, France}}

\ead[url]{http://www.quantware.ups-tlse.fr/dima}




\begin{abstract}
We propose a new Ising-PageRank model of opinion formation 
on a social network by introducing 
an Ising- or spin-like structure of the corresponding  Google matrix.
Each elector or node of the network 
has two components corresponding to a red or blue opinion in the society. 
Also each elector propagates either the red or the blue opinion
on the network so that the links between electors are described
by two by two matrices favoring one or the other of the two opinions. 
An elector votes for red or blue depending on the dominance of its red or blue 
PageRank vector components. We determine the dependence of the final 
society vote on the fraction of nodes with red (or blue) influence 
allowing to determine the transition for the election outcome border 
between the red or blue option. We show that this transition
border is significantly affected by the opinion 
of society elite electors composed of the top 
PageRank, CheiRank or 2DRank nodes of the network
even if the elite fraction is very small.
The analytical and numerical studies are 
preformed for the networks of
English Wikipedia 2017 and Oxford University 2006.
\end{abstract}

\maketitle


\section{Introduction}
\label{sec1}

The understanding of opinion formation in democratic societies is 
an outstanding challenge for scientific research \cite{zaller}.
In the last decade the development of social networks like
Facebook \cite{facebook}, Twitter \cite{twitter} and
VKONTAKTE \cite{vkontakte}, with  hundreds of millions of users,
demonstrated the growing
influence of these networks on social and political life.
Their growing influence on democratic elections is well 
recognized and highly debated \cite{guardian,nytimes}.
This makes the scientific analysis of opinion formation
on social networks of primary importance.  

The small-world and scale-free structures of social networks
(see e.g. \cite{dorogovtsev,vigna2012}),
combined with modern rapid communication facilities,  
leads to a rapid information propagation over 
networks of electors, consumers and citizens
generating their  instantaneous active reaction on social events.
This puts forward a request for new theoretical models
allowing to understand the 
opinion formation process in modern society.

Opinion formation was analyzed in the framework of various
interesting voter models described
in  detail in \cite{galam0,liggett,galamepl,watts2007,galam,
fortunatormp,krapivskybook,schmittmann}.
This research area became known as sociophysics 
\cite{galam0,galamepl,galam,fortunatormp,krapivskybook} for which 
a recent overview of various models is given in \cite{dong}.

Another type of model, called PageRank opinion formation (PROF) model,
was proposed in \cite{prof1,prof2,prof3}. In this model
each node of a directed network may have {\it red} or {\it blue}
opinion and the opinion of a each node is determined
by its neighboring nodes (on one link distance)
taken with the weight of PageRank probability
in the global network. Thus the PROF model takes into
account the PageRank concept
developed by Brin and Page \cite{brin}
which is now broadly used for various types of networks
(see reviews in \cite{meyer,rmp2015}).
This model leads to a number of interesting properties
of opinion formation for various examples of directed networks. 
However, a weak point of the PROF model
is that it assumes that the PageRank probabilities are known to the 
electors (nodes). This may be partially true since the
electors know approximately their social positions
in the society which can be assumed to be proportional 
to the PageRank probability.
But the exact global PageRank probabilities of neighbors
are most probably not known for a given local node.
Thus a new model  based on PageRank
properties and keeping the locality of 
knowledge about the network structure
is highly desirable.

With this aim we propose here a modified model,
called Ising-PageRank opinion formation model (Ising-PROF),
which corrects the above weak point of the PROF model
determining a more natural local process of opinion
formation still being based on the PageRank
concept. In this model an elector (node)
has two opinions (red or blue component)
being similar to a spin up or down state in the Ising model
\cite{ising,isinglandau}. A fraction $w_r$ of red oriented
nodes transfer their red influence via links to other nodes
while a fraction $w_b$ of blue oriented nodes 
propagates their blue influence $(w_r+w_b=1)$.
In this way the size of the Google matrix is doubled
since each node has now red and blue components (up or down 
states of an Ising spin). 
As a result the PageRank vector also has two components per node 
(of the original network) and its elector vote is determined by its largest 
PageRank components
(red or blue).  We assume that the top nodes of PageRank correspond 
to a political elite
of the social network whose opinion influences the opinions
of other members of the society \cite{zaller}.
Our results show that the elite influence,
related to the top PageRank electors, can significantly affect
the final vote on such a social network.

In our studies we consider as typical examples 
two types of real directed networks.
The first one is the English Wikipedia network of the year 2017 
with $N=5\, 416\, 537$ nodes and $N_l = 122\, 232\, 932$
links, studied recently in \cite{spgroups}, 
and the second one is the WWW network of 
Oxford University from the year 2006 
with $N= 200\,823$ nodes
and $N_l= 1\,831\,542$ links, studied in \cite{oxford}.

The paper is composed as follows: the Ising-PROF model is formally introduced 
in Sec. \ref{sec2}, numerical and analytical results for the model 
without elite are given in Sec. \ref{sec3}, numerical results for the elite 
influence are presented in Sec. \ref{sec4}, 
the polarization of opinion for individual nodes and 
the effect of resistance in opinion formation are studied 
in Secs. \ref{sec5}, \ref{sec6}.
The discussion of the results is presented in Sec. \ref{sec7}.

\section{Description of Ising-PageRank opinion formation  model}
\label{sec2}

We first remind the usual rules for the construction of the Google matrix $G$ 
from a given directed network with $N$ nodes and $N_l$ links 
described in detail in \cite{brin,meyer,rmp2015}
(we use here the notations of \cite{rmp2015}). 
For this one first defines 
the adjacency matrix $A_{ij}$ with elements $1$ if node (elector) $j$ 
points to  node (elector) $i$ and zero otherwise. 
In this case, the elements of the Google matrix take the standard form 
$G_{ij} = \alpha S_{ij} + (1-\alpha)\, v(i)$ 
\cite{brin,meyer,rmp2015},
where $S$ is the matrix of Markov transitions with elements  
$S_{ij}=A_{ij}/d_j$, 
$d_j=\sum_{i=1}^{N}A_{ij}\neq 0$ being the node $j$ out-degree
(number of outgoing links from node $j$) and with $S_{ij}=v(i)$ 
if $j$ has no outgoing links (dangling node). Here the vector $v$ 
(with $\sum_i v(i)=1$ and $v(i)\ge 0$) is also called 
personalization or teleportation vector \cite{brin,meyer}. 
Furthermore the parameter 
$0< \alpha <1$ is the damping factor which for a random surfer
determines the probability $(1-\alpha)$ to jump (or ``teleport'') 
to any node $i$ (with 
relative weight $v(i)$). The usual standard values are $v(j)=1/N$ and 
$\alpha=0.85$. For the teleportation vector it is possible to choose 
a different vector and one may also choose two different vectors for 
the dangling node columns of $S$ and the columns of the contribution 
proportional to $(1-\alpha)$. 

The PageRank is the right eigenvector of the Google matrix
($G P = \lambda P, \lambda=1$) of the largest eigenvalue $\lambda=1$. 
It has positive components $P(j)$ normalized to unity
($\sum_j P(j)=1$). We note that the largest unit eigenvalue is not 
degenerate for $\alpha<1$ and the PageRank can be efficiently computed 
from the power iteration method with a convergence rate $\sim \alpha^t$ (with 
$t$ being the iteration time).

\begin{figure}[h]
\begin{center}
\includegraphics[width=0.44\textwidth]{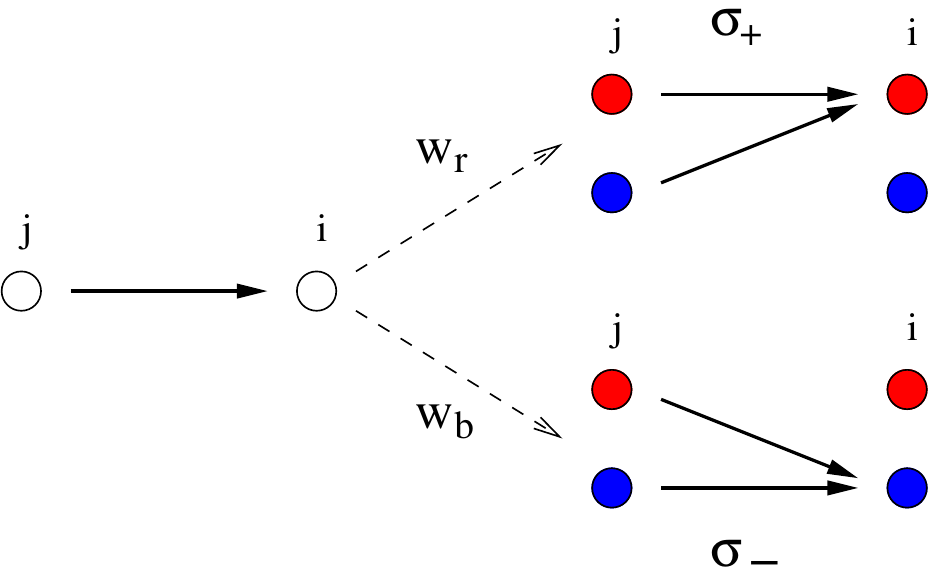}
\end{center}
\caption{\label{fig1} 
Schematic description of the construction of the Ising-network from 
a given directed network. Each node of the original network is doubled 
in a red and blue node and gets either (with probability $w_r$) a preference 
to point to other red nodes or (with probability $w_b=1-w_r$) 
a preference to point to other blue nodes. 
Each link $j \to i$ of the original network 
is replaced by the two links $j_{\rm red}\to i_{\rm red}$ and 
$j_{\rm blue}\to i_{\rm red}$ (if $j$ has a red preference) 
or the two links
$j_{\rm red}\to i_{\rm blue}$ and $j_{\rm blue}\to i_{\rm blue}$
(if $j$ has a blue preference); 
$j_{\rm red}$ ($j_{\rm blue}$) designate the index of the red or blue 
node of the Ising-network with $j$ being the node index of the 
original network. 
}
\end{figure}

We now introduce the Google matrix for the Ising-PageRank opinion 
formation model (Ising-PROF). First, each node of the original network is 
doubled into a pair of red and blue nodes giving a total network size of $2N$.
Furthermore, we attribute randomly to each node (of the original network) 
either a vote preference for red with probability $w_r$ or blue 
with probability $w_b=1-w_r$ where $0\le w_r\le 1$ is a global parameter 
for the overall vote preference. Therefore for each random realization 
there is approximately a fraction $w_r$ of nodes with red preference and 
a fraction $w_b$ with blue preference. The links are 
also doubled: for each link from a node $j$ to $i$ of the original network 
we will have two links from both $j$ nodes (blue and red) 
to the red node $i$ if $j$ has a preference for red or to the blue node 
of $i$ if $j$ has a preference for blue. This scheme is also illustrated 
in Fig.~\ref{fig1} and mathematically it implies that in the (original) 
adjacency matrix each unit entry $A_{ij}$ is replaced either by a certain 
$2 \times 2$ matrix $\sigma_+$ if $j$ has a red preference or 
by another $2 \times 2$ matrix $\sigma_-$ if $j$ has a blue preference 
where the $2 \times 2$ matrices $\sigma_\pm$ are given by:
\begin{equation}
\label{eq_sigmadef}
\sigma_+=\left(\begin{array}{cc}
1 & 1 \\
0 & 0 \\
\end{array}\right)
\quad,\quad
\sigma_-=\left(\begin{array}{cc}
0 & 0 \\
1 & 1 \\
\end{array}\right)\ .
\end{equation}
 
This provides a larger $2N\times 2N$ adjacency matrix $A_2$ 
from which we construct the $2N\times 2N$ Google matrix, 
noted as $G_2$ in the usual way as described above. 
However, we choose 
a particular teleportation vector $v_r(i)=w_r/N$ ($v_b(i)=w_b/N$) for 
the red (blue) component $v_r(i)$ ($v_b(i)$) (instead of the uniform 
choice $1/(2N)$ for both components).

The PageRank vector $P$ of $G_2$ (defined by 
$G_2 P = P$) has red (blue) components 
$P_r(i)$ ($P_b(i)$) where $i$ belongs to the set of original nodes 
and the sum normalization reads $\sum_i [P_r(i)+P_b(i)]=1$. 
In this work we study in particular two quantities 
derived from this PageRank vector which is the total 
PageRank probability for red (or the partial PageRank norm for red nodes)
 given by~:
\begin{equation}
\label{eq_Prdef}
P_r=\sum_{i=1}^N P_r(i)
\end{equation}
and the total vote for red given by 
\begin{eqnarray}
\label{eq_Vrdef}
V_r&=&\frac{1}{N}\#\Bigl\{\mbox{nodes }i\mbox{ with }P_r(i)>P_b(i)\Bigr\}\\
\nonumber
&&+\frac{1}{2N}\#\Bigl\{\mbox{nodes }i\mbox{ with }P_r(i)=P_b(i)\Bigr\}
\end{eqnarray}
which is the fraction of nodes $i$ such that 
$P_r(i)>P_b(i)$ (rare cases of $P_r(i)=P_b(i)$ count with a relative weight 
of $1/2$).
The complementary vote for blue is given by $V_b=1-V_r$. 
The red opinion wins the global society vote if 
the sum over all red votes of electors is larger than 50\%.

\section{Analytical results and estimates}
\label{sec3}

As in the above section we denote by $P_r(i)$ and $P_b(i)$ the PageRank 
components for red or blue nodes of $G_2$. Furthermore, we denote by $P(i)$ 
the PageRank vector of the Google matrix $G$ of the 
original network with $N$ nodes. 
Furthermore let 
$\tilde P(i)=P_r(i)+P_b(i)$. We first show that for our model we have exactly 
$\tilde P(i)=P(i)$. 
The PageRank equation of $G_2$ for red nodes reads:
\begin{equation}
\label{eq_PR1}
P_r(i)=\alpha\sum_{j\in L_i} \frac{n_j}{d_j}\tilde P(j)+
\alpha \frac{w_r}{N}\sum_{j\in D} \tilde P(j)
+(1-\alpha)\frac{w_r}{N}\sum_{j=1}^N \tilde P(j)
\end{equation}
where  $L_i$ is the set of nodes $j$ such 
that there is a link $j\to i$ (this set may be empty), $D$ is the set 
of dangling nodes, 
$d_j$ is the outdegree of node $j$ being the number nodes $k$ such there is 
a link $j\to k$. All these quantities 
refer to the original network. Furthermore, $w_r$ is the overall vote 
preference for red introduced in the last section. 
$n_j$ is a random number being either $1$ (with probability $w_r$) 
for nodes $j$ with red preference or $0$ (with probability 
$w_b=1-w_r$) for nodes $j$ with blue preference. The average and 
variance of $n_j$ are obviously given by~:
\begin{equation}
\label{eq_simple_av}
\langle n_j\rangle=w_r\quad,\quad 
\langle\delta n_j^2\rangle=\langle n_j^2\rangle-\langle n_j\rangle^2
=w_r(1-w_r)
\end{equation}
since $n_j^2=n_j$. We also have $\langle \delta n_j\delta n_k\rangle=0$ 
if $j\neq k$. The second and third 
sum terms in (\ref{eq_PR1}) take into account our particular choice 
for the teleportation vector for the Ising-PROF model introduced in the 
last section. 

\begin{figure}[h]
\begin{center}
\includegraphics[width=0.48\textwidth]{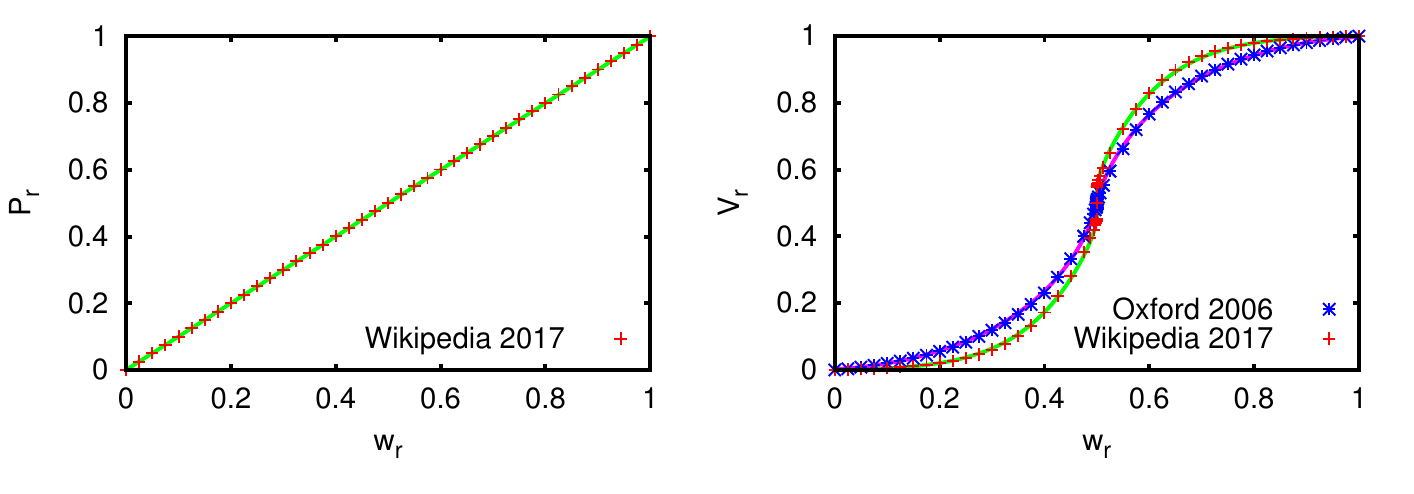}
\end{center}
\caption{\label{fig2} 
{\em Left panel:}
Total PageRank probability $P_r$ for red 
nodes depending on $w_r$ for English Wikipedia of 2017; the straight line 
corresponds to the theoretical expression $P_r=w_r$. The data for Oxford 2006, 
not shown, are on graphical precision identical to the data of Wikipedia of 
2017. 
{\em Right panel:} The vote quantity $V_r$ given as the fraction of nodes 
where $P_r(i)>P_b(i)$ depending on $w_r$ 
for English Wikipedia of 2017 and Oxford 2006. The full lines correspond 
to the rescaled expression $V_r^{\rm (th)}\bigl(5/6(w_r-0.5)+0.5\bigr)$ 
where 
$V_r^{\rm (th)}(w_r)$ is the theoretical expression (\ref{eq_Vrth}) based 
on the assumption Gaussian distributed $P_r(i)$. 
All discrete data points in this figure (and in all subsequent figures 
except Fig.~\ref{fig6}) were obtained from an 
ensemble average over 10 different realizations of different attributions of 
$\sigma+$ or $\sigma_-$ for each node $i$ and the resulting statistical 
error bars are below $10^{-3}$ (below size of data points) 
for both quantities $P_r$ and $V_r$. 
}
\end{figure}

The equation for $P_b(i)$ is similar with the replacement $n_j\to 1-n_j$ 
and $w_r\to 1-w_r$. We note that on the right hand side only the sum
$\tilde P(j)=P_r(j)+P_b(j)$ appears
due to the structure of $\sigma_{\pm}$. 
If we add the equations for $P_r(i)$ and $P_b(i)$ we obtain for 
$\tilde P(i)$ the exact PageRank equation of the original network such that 
exactly $\tilde P(i)=P(i)$ and $\tilde P(i)$ is no longer random which gives 
a great simplification. We have also numerically verified that this 
property holds up to numerical precision ($10^{-13}$). 

Using (\ref{eq_simple_av}) we can analytically compute the ensemble average 
of (\ref{eq_PR1}) which gives 
$\langle P_r(i)\rangle=w_r P(i)$ and therefore we obtain 
exactly $\langle P_r\rangle =\sum_i \langle P_r(i)\rangle =w_r$ which 
is numerically clearly confirmed by the left panel 
of Fig.~\ref{fig2}. 

Furthermore, $P_r(i)$ is a sum of random variables $n_j$ (with some 
coefficients). If we assume that there are many terms (if $\#L_i\gg 1$, i.e. 
many incoming links) then the central limit theorem implies that $P_r(i)$
is approximately Gaussian distributed (however, in realistic networks 
with modest numbers in the sets $L_i$ this is probably not very exact).
Also the variance of $P_r(i)$ can be computed from (\ref{eq_PR1}) and 
(\ref{eq_simple_av}):
\begin{equation}
\label{eq_variance1}
\langle \delta P_r(i)^2\rangle=
\alpha^2 w_r(1-w_r)\sum_{j\in L_i} \frac{P(j)^2}{d_j^2}\ .
\end{equation}
If the assumption of $P_r(i)$ being a Gaussian variable is valid the known 
mean $\langle P_r(i)\rangle = w_r\,P(i)$ and variance (\ref{eq_variance1}) 
are sufficient to characterize the full distribution $p_{\rm gauss}(P_r(i))$. 
The node $i$ contributes to a red vote 
if $P_r(i)>P_b(i)=P(i)-P_r(i)\ \Leftrightarrow\ P_r(i)>P(i)/2$.
Therefore the probability $V_r(i)$ of a red vote of node $i$ can be obtained 
as 
\begin{equation}
\label{eq_vote1}
V_r(i)=\int_{P(i)/2}^\infty dP_r(i)\ p_{\rm gauss}(P_r(i))
\end{equation}
which gives with the help of (\ref{eq_variance1}) and the average 
$\langle P_r(i)\rangle=w_r P(i)$:
\begin{equation}
\label{eq_vote2}
V_r(i)=\frac{1}{2}\left(1-
\mbox{erf}\left(\frac{0.5-w_r}{\alpha\,a_i\,\sqrt{w_r(1-w_r)}}
\right)\right)
\end{equation}
where
\begin{equation}
\label{eq_adef}
a_i=\frac{1}{P(i)}\sqrt{2\sum_{j\in L_i} \frac{P(j)^2}{d_j^2}}
\end{equation}
is a quantity that can be efficiently computed (for all nodes $i$ 
simultaneously). We remind that in (\ref{eq_variance1}) and (\ref{eq_vote2}) 
the parameter $\alpha=0.85$ is the damping factor. 
Note that it is possible that $L_i$ is an empty set 
(if row $i$ of the adjacency matrix is empty, i.e. if node $i$ is a 
dangling node for $G^*$). In this case 
$a_i=0$ and in (\ref{eq_vote2}) we obtain the Heaviside function 
$V_r(i)=H(w_r-0.5)$ which is not a problem. It turns out that for 
English Wikipedia 2017 and Oxford 2006 the quantity $a_i$ has a 
maximal value of about $1.66$ and is typically between $0.2$ and $1$ 
for most nodes. The average and variance (with respect to the node index) 
of $a_i$ are $\langle a_i\rangle=0.338$, $\langle\delta (a_i)^2\rangle=0.062$ 
($\langle a_i\rangle=0.523$, $\langle\delta (a_i)^2\rangle=0.121$)
and there is 
also a finite fraction of nodes with $a_i=0$ which is $9.68\times 10^{-2}$ 
($2.14\times 10^{4}$) for Wikipedia 2017 (Oxford 2006). 

Averaging (\ref{eq_vote2}) with respect to all nodes gives the theoretical 
expression for the total vote:
\begin{equation}
\label{eq_Vrth}
V_r^{\rm (th)}(w_r)=\frac{1}{N}\sum_i V_r(i)
\end{equation}
which can be computed numerically with a modest effort. The expression 
(\ref{eq_vote2}) corresponds roughly to a smoothed step function 
with $V_r(i)$ being $0$ (or $1$) for $w_r=0$ (or $w_r=1$) and a nonlinear 
shape such that the slope at $w_r=0.5$ is proportional to the parameter 
$a_i^{-1}$. Even though the value of $a_i$ depends on the node index $i$ 
the total vote (\ref{eq_Vrth}) has a similar nonlinear shape close 
to a smoothed step function. However, due to the small but finite fraction 
about $0.1$ ($0.0002$) for Wikipedia (Oxford) of nodes with $a_i=0$ 
(i.e. nodes with empty sets $L_i$) 
there is a small vertical finite step (with infinite slope) in the 
curve of $V_r$ versus $w_r$ at $w_r=0.5$. The vertical size of this step 
corresponds exactly to this fraction. 

The overall shape and also the small vertical finite step are 
confirmed by the numerical data visible in the right panel of 
Fig.~\ref{fig2} for Wikipedia 2017 and the WWW-network of Oxford 2006. 
However there is not a perfect agreement of (\ref{eq_Vrth}) 
with the numerical data but if we apply a slight 
rescaling by using $V_r^{\rm (th)}\bigl(5/6(w_r-0.5)+0.5\bigr)$ 
(instead of $V_r^{\rm (th)}(w_r)$) there is a very good matching with 
the numerical data. It seems that the Gaussian assumption underestimates 
slightly the probability of having $P_r(i)$ values far from its average 
$w_r P(i)$. Most likely the number of terms in 
the set $L_i$ is too small for many nodes $i$ such that there is 
not a perfect justification for the use of the central limit theorem. 
Since the distribution of each $n_i$
has only two values we have indeed to add very many terms to obtain a nice 
Gaussian. Furthermore, the coefficients ($P(j)/d_j$) also fluctuate 
with the node index $j$ such that even less terms contribute effectively 
in the sum of random variables. 

One can also try the expression (\ref{eq_vote2}) as fit expression 
for the numerical data of the total vote (using $a_i$ as fit parameter). 
It turns out that this does not work very well. However if we add 
two such functions (with three 
fit parameters: two $a_i$ values and the weight between both terms) there 
is a quite good (but not really perfect) fit.

\section{Results for elite influence in Ising-PROF model}
\label{sec4}

The results presented above are rather natural and bring no surprise.
However, the  Ising-PROF model introduced in Sec. \ref{sec2} 
is local and thus has advantages
in comparison to the PROF model proposed in \cite{prof1}.
In particular, it can be generalized to study the influence 
of elite opinion on the final vote. 
We select three types of elite on our social network based on different 
rankings. 
For the first ranking type all nodes are ordered in 
decreasing order of PageRank probability (of the original network) 
noted by the index $K(j)$ ($1\leq K(j) \leq N$) with the highest probability
$P(j)$ if $K(j)=1$ and smallest probability at
$P(j)$ if $K(j)=N$.
Thus the nodes $j$ with $K(j)=1,2,3...$ are considered as the most influential
electors (nodes) corresponding to party leaders, government members etc. 
The second type of elite is determined from
the CheiRank probabilities $P^*(j)$ (of the original network)
giving the ordering index $K^*(j)$. 
The CheiRank vector is the PageRank vector of the original network
with inverted direction of all links
(see detailed description in \cite{cheirank,zzswiki,rmp2015}).
While the PageRank probability is on average 
proportional to the number of ingoing links
the CheiRank probability is on average 
proportional to the number of outgoing links.
In a certain sense we can consider the top
CheiRank electors $j$ with $K^*(j)=1,2,3...$
to be analogous to press and television.
The third type of elite is given by the top nodes of
2DRank which represents a combination of
PageRank and CheiRank top nodes $j$ with index $K_2(j)=1,2,...N$
(see description in \cite{rmp2015,zzswiki}).

\begin{figure}[h]
\begin{center}
\includegraphics[width=0.48\textwidth]{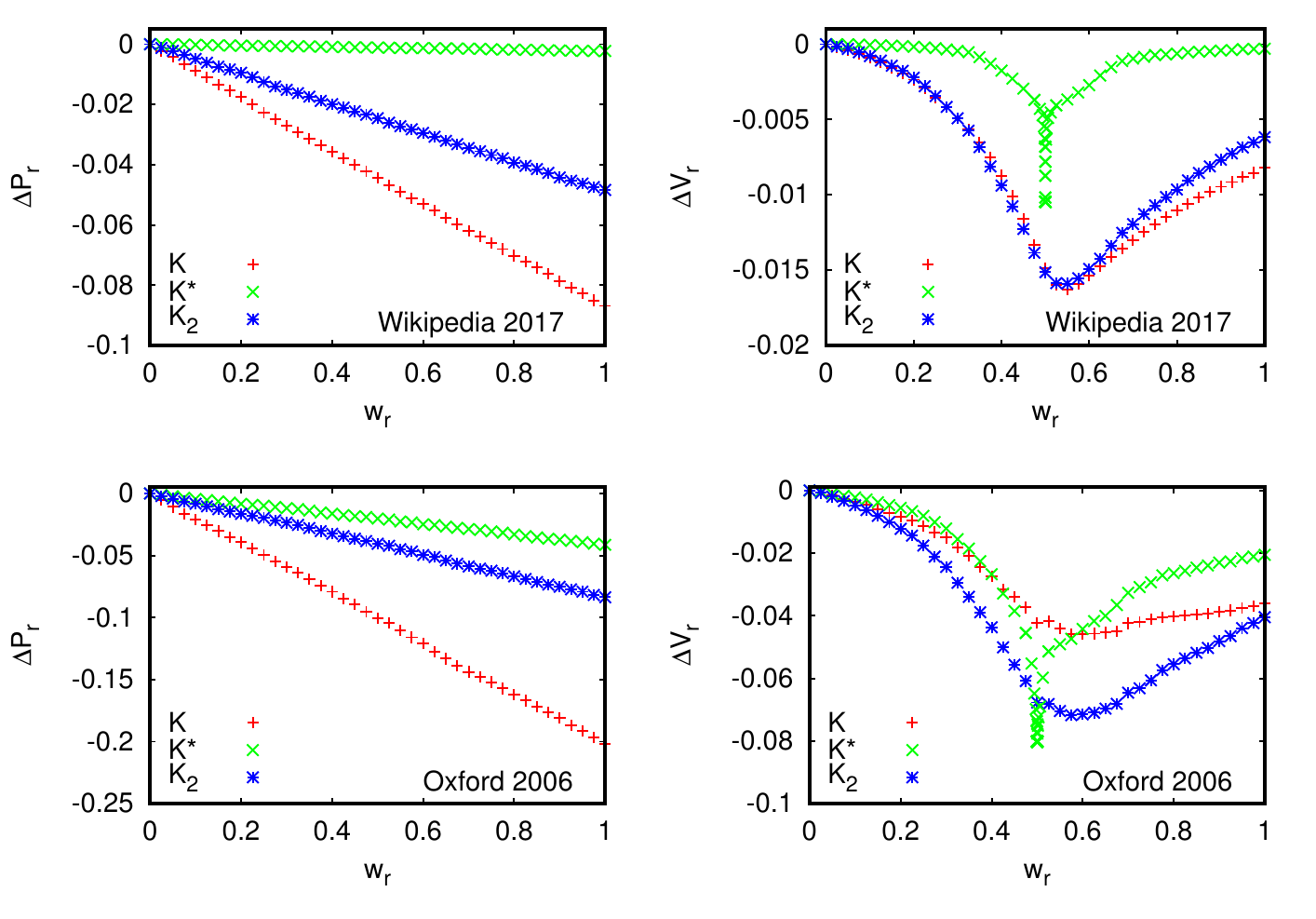}
\end{center}
\caption{\label{fig3} 
Dependence of elite induced variation of red PageRank 
probability $\Delta P_r$ and red vote $\Delta V_r$ 
on fraction $w_r$ of red nodes in the whole network. 
The variations
$\Delta P_r=P_{r,\rm el}-P_r$ (left panels) and 
$\Delta V_r=V_{r,\rm el}-V_r$ (right panels) are shown 
in dependence of $w_r$ where 
$P_{r,\rm el}$ and $V_{r,\rm el}$ are obtained from a model of 
$N_{\rm el}=1000$ elite nodes with $w_{r,\rm el}=0$ while the other nodes 
correspond the probability $w_r$. Top (bottom) panels correspond 
to English Wikipedia 2017 (Oxford 2006). In each panel the three different 
type of data points correspond to the cases where the elite nodes 
are obtained as the top $1000$ nodes according to $K$-rank (red plus 
symbols), $K^*$-rank (green crosses) or $K_2$-rank (blue stars). 
}
\end{figure}

To determine the influence of elite on the society final
vote we modify the model of Sec. \ref{sec2} such that for 
$N_{\rm el}$ elite notes $j$ with 
$1 \leq K(j),K^*(j),K_2(j) \leq N_{\rm el}$ 
the probability of vote preference for 
red is modified to $w_{r,\rm el}$ which is different from $w_r$ which 
applies to the remaining nodes. (We keep however, 
since the elite fraction is very small, the same values 
$v_r(i)=w_r/N$ and $v_b(i)=w_b/N$ for {\em all} nodes $i$ for the 
teleportation vector as in the initial uniform Ising-PROF model.)
Therefore $w_{r,\rm el}$ will be the 
approximate fraction of red nodes in the set of elite nodes while $w_r$ 
is the fraction of red nodes in the set of remaining nodes. 
We consider for $w_{r,\rm el}$ values between $0$ and $0.5$ 
and $w_r$ values between $0$ and $1$ 
(since red and blue can be interchanged there is no reason
to consider $w_{r,\rm el} >0.5$).
Thus for $w_{r,\rm el}=0$ all elite nodes belong to
the blue fraction ($w_{b,\rm el}=1-w_{r,\rm el}=1$). 
Usually we consider $N_{\rm el}\ll N$ 
so that these elite nodes should not affect 
the global vote $V_r$ if they 
were randomly and homogeneously
distributed over the whole network of $N$ nodes.
But we show that this small fraction 
distributed only over elite electors
significantly affects the final $V_r$ vote. 
To characterize the influence of elite
we introduce the variation of the total PageRank
probability on red nodes $\Delta P_r =P_{r,\rm el} -P_r$
induced by elite and respectively 
the variation of the global red vote
$\Delta V_r = V_{r,\rm el} -V_r$ where $P_r$ and $V_r$ are obtained from 
the Ising-PROF model without elite for which analytical and numerical results 
were given in the last section. 

In principle, the analytical argument for $\tilde P(i)=P(i)$ also holds 
for the case of elite nodes and we can also try to compute the 
average and variance of $P_r(i)$ which requires 
in (\ref{eq_simple_av}) to replace $w_r$ by $w_j$ where $w_j$ now depends 
on the node $j$ and takes either the value $w_{r,\rm el}$ 
if $j$ is an elite node or $w_r$ otherwise. The resulting expressions are 
therefore more complicated and depend more strongly on the particular 
network structure and also on the type of elite nodes chosen. Therefore 
they do not allow a simple evaluation and in this section will we 
concentrate on the numerical results. 

\begin{figure}[h]
\begin{center}
\includegraphics[width=0.48\textwidth]{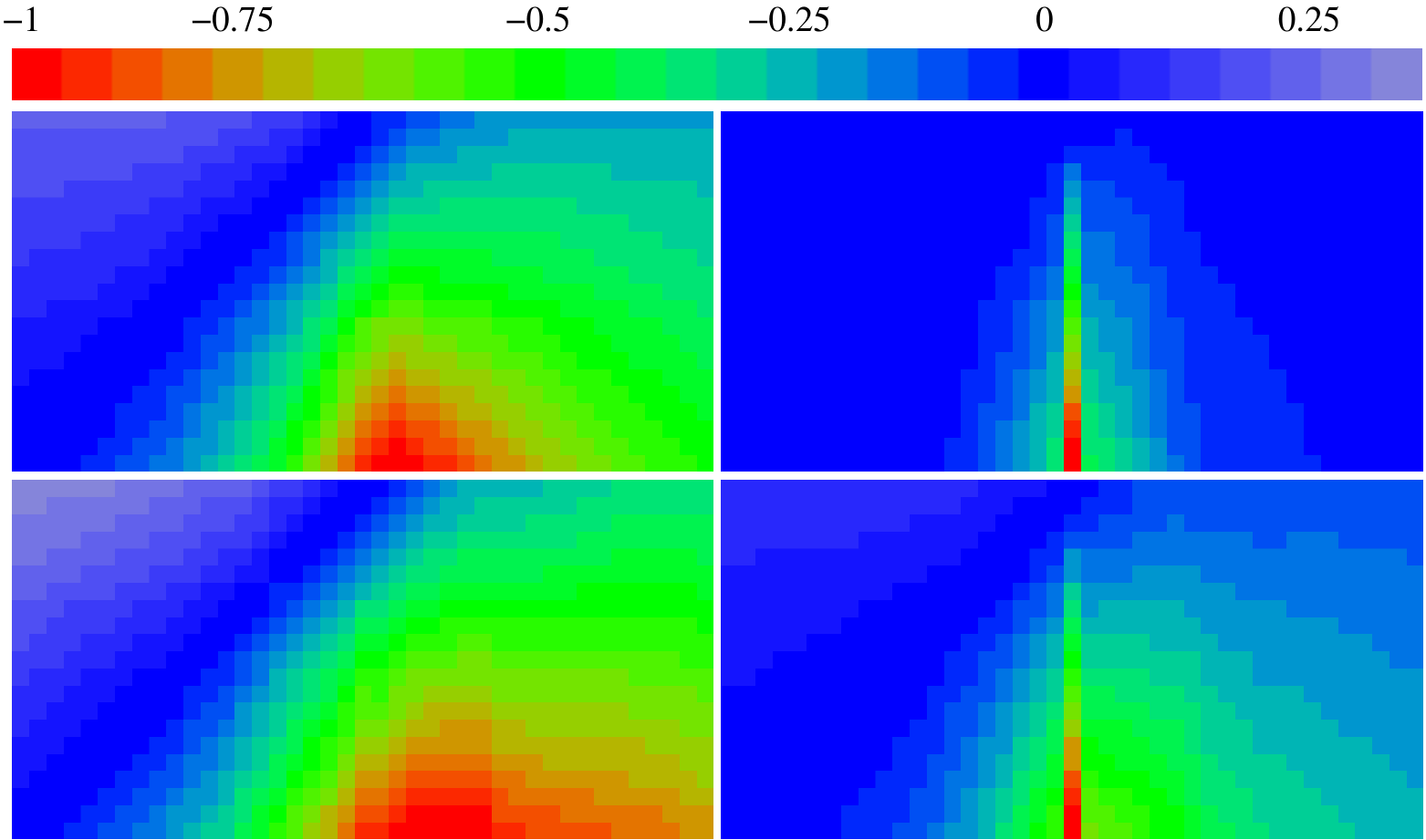}
\end{center}
\caption{\label{fig4} 
Elite induced variation of red vote $\Delta V_r=V_{r,\rm elite}-V_r$
is shown by color for 
different values of $w_r$ (corresponding to horizontal axis 
with $w_r\in[0,1]$) and of $w_{r,\rm el}$ 
(corresponding to vertical axis with $w_{r,\rm el}\in[0,0.5]$). 
The numerical values of the top color bar correspond to the 
fraction $\Delta V_r/V_{\rm max}$ with 
$V_{\rm max}$ being the maximum of $|\Delta V_r|$ in the range 
of considered $w_r$ and $w_{r,\rm el}$ values.
Left (right) panels correspond to elite nodes given 
as the top $N_{\rm el}=1000$ nodes 
for PageRank index $K$ (CheiRank index $K^*$). Top (bottom) panels correspond 
to English Wikipedia 2017 (Oxford 2006). 
The values of $V_{\rm max}$ for each panel 
are $0.016$ (top left), $0.011$ (top right), 
$0.046$ (bottom left), $0.080$ (bottom right). 
}
\end{figure}

The dependence of 
$\Delta P_r$ and $\Delta V_r$ on $w_r$ is shown in Fig.~\ref{fig3}
for Wikipedia 2017 and the WWW-network of Oxford 2006 
for the case when all $N_{\rm el}=1000$ nodes of elite
have a blue preference $w_{r,\rm el}=0$ (all three types of elite are shown).
Here we have  $N_{\rm el} \ll N$ so that 
a random distribution of these $N_{\rm el}=1000$ nodes
over the whole network
gives a negligible variation
of $\Delta P_r$ and $\delta V_r$.
However, when $N_{\rm el}$ occupies the top
rank positions of $K, K^*, K_2$
we obtain significant changes
of  $\Delta P_r$ and $\Delta V_r$.
The dependence of $\Delta P_r$
on $w_r$ remains approximately linear
but the red component probability
is reduced in comparison to the $P_r$ value
in elite absence (see Fig.~\ref{fig2} left panel).
The change of red vote $\Delta V_r$
has a rather nontrivial dependence on $w_r$
with a maximum absolute value 
being about $0.016$ for Wikipedia 
and $0.075$ for Oxford networks.
For the critical point with $w_r \approx 0.5$
the blue elite induces a vote gain for the 
blue party with about an 1.5\% advantage
for Wikipedia PageRank or 2DRank elite 
and a 7.5\% advantage for Oxford PageRank elite 
(4\% for 2DRank elite). 
The cases of PageRank and 2DRank elite
have a smooth dependence 
$\Delta V_r(w_r)$ while for the 
CheiRank elite 
this dependence is significantly peaked
near $w_r \approx 0.5$.
For the Wikipedia case the behavior of $\Delta V_r(w_r)$
is rather similar between PageRank and 2DRank elite cases
while the CheiRank elite produces a smaller change of vote.
For the Oxford network the situation is a bit different:
the CheiRank elite gives a bit stronger
variation of the vote being strongly peaked
near $w_r \approx 0.5$,
the 2DRank elite gives slightly smaller
changes of the vote as compared to the PageRank elite
with a factor of about $0.7$ between the maximal amplitudes 
for both (at $w_r\approx 0.6$). 

This shows that the network structure
plays a certain role in the 
elite vote influence
even if the difference between the three elite types
is only about about 30-40\%.
Of course, in the case of Oxford the 
fraction of elite nodes is
larger than for Wikipedia
($N_{\rm el}/N \approx 1/200$ and $1/5000$ respectively)
and due to this the change of vote
$\Delta V_r$ is larger for Oxford. 
We investigate the dependence on $N_{\rm el}/N$ below.  

\begin{figure}[h]
\begin{center}
\includegraphics[width=0.48\textwidth]{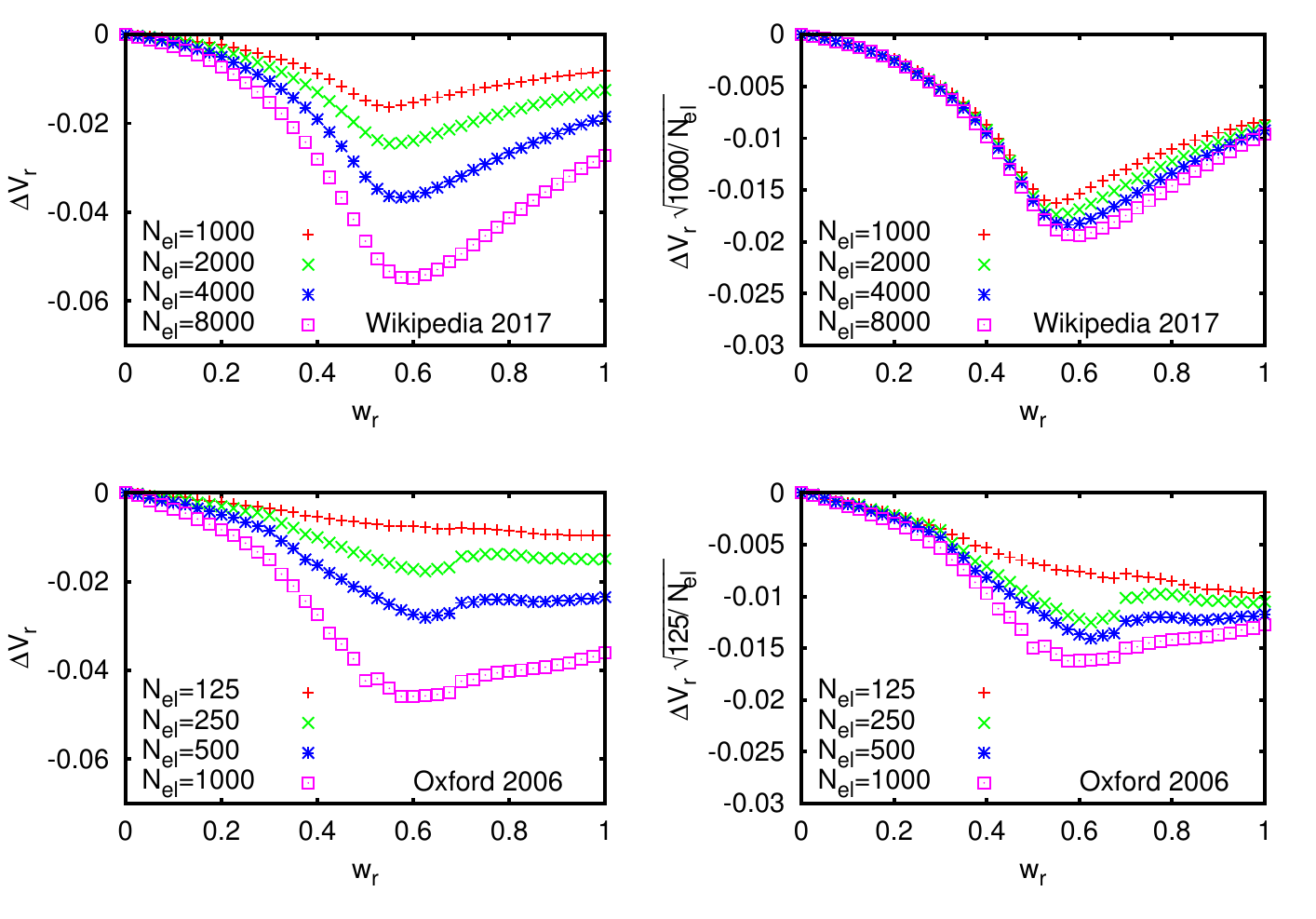}
\end{center}
\caption{\label{fig5} The dependence of 
$\Delta V_r=V_{r,\rm el}-V_r$ on $w_r$ for $w_{r,\rm elite}=0$ and 
for various values of elite nodes $N_{\rm el}$ obtained as top $N_{\rm el}$ 
nodes from $K$ rank. Top (bottom) panels correspond 
to English Wikipedia 2017 (Oxford 2006). 
Left panels show directly $\Delta V_r$ versus $w_r$ and right panels show 
the rescaled quantity $\Delta V_r \sqrt{1000/N_{\rm el}}$ (top) 
or $\Delta V_r \sqrt{125/N_{\rm el}}$) (bottom) versus $w_r$ indicating 
an approximate dependence $\Delta V_r\sim\sqrt{N_{\rm el}}$ for sufficiently 
small values of $w_r$. 
}

\end{figure}

In Fig.~\ref{fig3} we considered the case when all elite nodes have 
blue vote preference, i.e. $w_{r,\rm el}=0$. 
The variation of $\Delta V_r$ with $w_{r,\rm el}$
is shown in Fig.~\ref{fig4}. We see that for the PageRank elite the
variation of red vote $\Delta V_r$ being close to its maximum
value of about $1.5$\% can be reached also at $w_{r,\rm el} \approx 0.25$. 
For larger values $w_{r,\rm el} >0.25$ the variation  $\Delta V_r$
approaches zero at $w_{r,\rm el} =0.5$. For the case of CheiRank elite
the distribution of the variation of $\Delta V_r$
is mainly concentrated in a vicinity of the critical
probability $w_r \approx 0.5$ in agreement with the peaked minima 
visible in Fig.~\ref{fig3}. 

We note that $\Delta V_r$ may also have 
positive values in the region $w_{r,\rm el}>w_r$ (top left triangle in 
the panels of Fig.~\ref{fig4}) since in this case 
nodes with red preference in the elite fraction
increase a bit the global red vote. However, in this region
the red vote is small and this variation
does not play an important role.

The dependence of $\Delta V_r$ on $N_{\rm el}$ is shown 
in Fig.~\ref{fig5} for  $w_r =0$. The are well described by
a square-root dependence $\Delta V_r \propto \sqrt{N_{\rm el}/N}$ 
for sufficiently small values of $w_r$. 
To be more precise, from our numerical data in the vicinity of 
$w_r \approx 0.5$ we obtain the dependence
\begin{equation}
\label{eq_dVrdep}
\Delta V_r = -B(1-2w_{r,\rm el})\sqrt{N_{\rm el}/N}
\end{equation}
with a numerical constant $B\approx 1.114\pm 0.003$
for Wikipedia and $B \approx 0.611\pm 0.003$ for Oxford
in the case of PageRank elite. 
For 2DRank (CheiRank) elite we have approximately 
$B \approx 1.116\pm 0.003$ ($B \approx 0.773\pm 0.002$) 
for Wikipedia and 
$B \approx 0.960\pm 0.002$ ($B \approx 1.145\pm 0.003$) 
for Oxford. 
The numerical values of $B$ were obtained from a fit at $N_{\rm el}=1000$. 
For Wikipedia it also applies to other values of $N_{\rm el}$ as can 
be seen in the top right panel of Fig.~\ref{fig5} confirming 
the above square-root dependence of $\Delta V_r$ also at $w_r=0.5$. 
For Oxford there are at $w_r=0.5$ already visible modest deviations 
(see bottom right panel of Fig.~\ref{fig5}). However, here 
the square-root dependence is still rather correct for $w_r<0.2$. 

We explain the square-root dependence by the fact of 
diffusive accumulation of fluctuations,
like in the central limit theorem,
as discussed in equations (\ref{eq_vote1})-(\ref{eq_adef}).
However, an exact analytic derivation of the dependence
(\ref{eq_dVrdep}) still needs to be obtained.

\section{Polarization of opinion for individual nodes}
\label{sec5}

\begin{figure}[h]
\begin{center}
\includegraphics[width=0.48\textwidth]{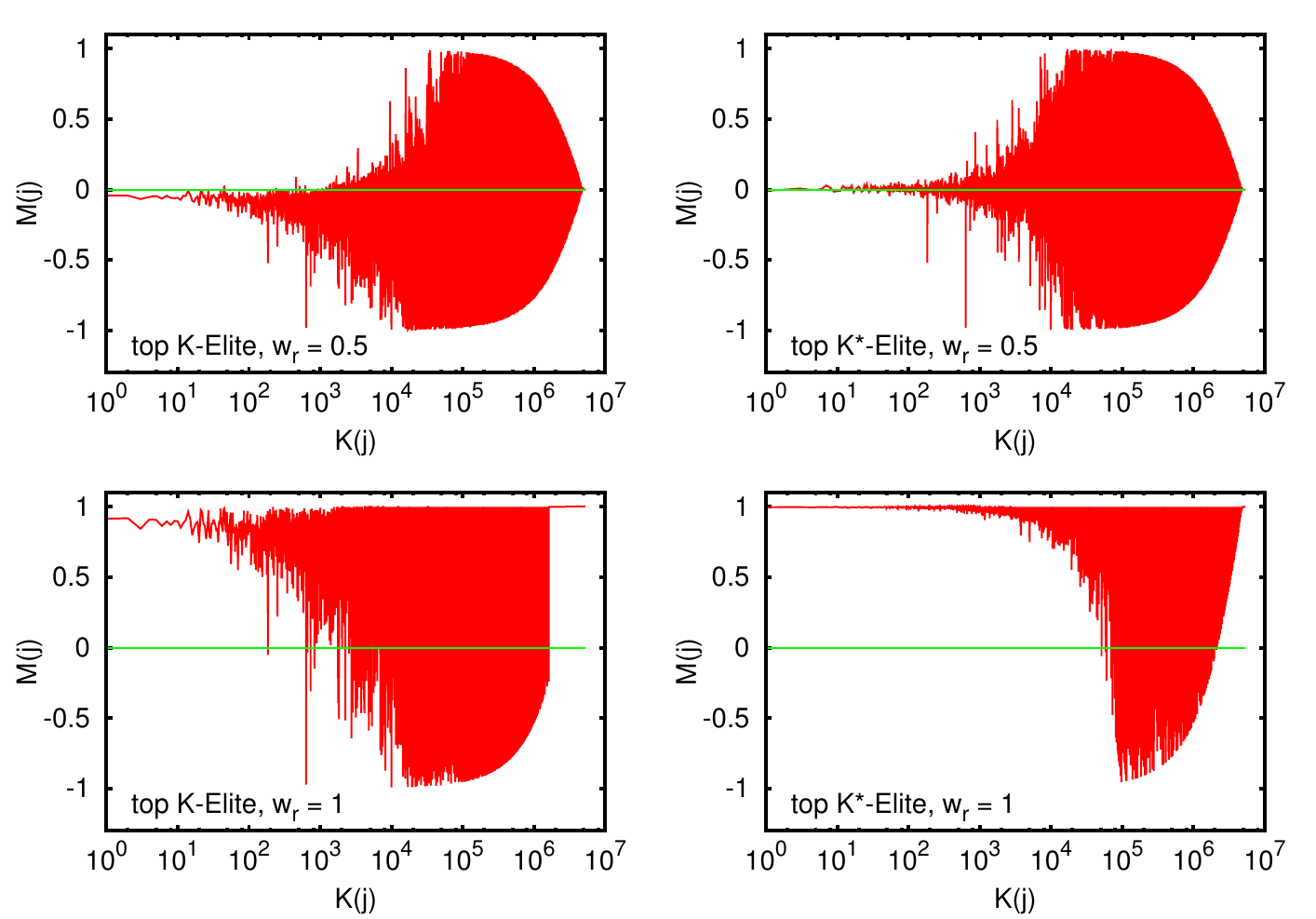}
\end{center}
\caption{\label{fig6} Dependence 
of $M(j)=(P_r(j)-P_b(j))/(P_r(j)+P_b(j))$ on rank $K$ index of node $j$ 
for $w_{r,\rm el}=0$ of top 1000 rank nodes 
for English Wikipedia 2017 (all panels) and for one individual random 
realization of attribution of $\sigma_\pm$ matrices to nodes. 
Top (bottom) panels correspond to $w_r=0.5$ ($w_r=1$). 
Left (right) panels correspond to elite nodes as top 1000 nodes 
from PageRank $K$ index (CheiRank $K^*$ index).
The green line shows zero polarization;
the horizontal axis shows the PageRank index $K$ (of the original 
network) in log scale for {\em all} panels. 
}
\end{figure}

It is interesting to analyze the polarization of individual nodes
in presence of elite influence.
For this we determine the polarization of a node $j$ defined as
\begin{equation}
\label{eq_Mdef}
M(j)=\frac{P_r(j)-P_b(j)}{P_r(j)+P_b(j)}\ .
\end{equation}
The influence of elite (with parameters $w_{r,\rm el}=0$, $N_{\rm el}=1000$) 
for Wikipedia on this polarization is shown in Fig.~\ref{fig6} for 
$w_r=0.5$ (top panels) and $w_r=1$ (bottom panels) 
with PageRank elite (left panels) or CheiRank elite (right panels). 

In all four cases the typical value of the polarization $M$ for the first 
top PageRank nodes (with $K(j)$ below $10^2$ for PageRank elite or 
below $10^3$ for CheiRank elite) are rather close to the 
ideal values $M\approx 0$ for $w_r=0.5$ or $M\approx 1$ for $w_r=1$ with 
only weak fluctuations. For larger values of $K(j)$ the value of $M$ 
strongly fluctuates between $-1$ and $1$. 

However, for $w_r=0.5$ and PageRank elite the top PageRank nodes 
still remain mostly blue but only with a weak polarization 
$M \approx -0.1$ (there are only 8 PageRank elite nodes
which change the polarization from blue to red) 
while the value $w_{r,\rm el}=0$ should suggest $M\approx -1$ 
for these elite nodes. Apparently the influence of the bulk value 
$w_r=0.5$ from the other nodes reduces strongly the polarization of the 
top PageRank (or elite) nodes but is not sufficient to change the sign. 

For the case of CheiRank elite 
the elite nodes do not coincide with the 
top PageRank nodes and their positions are quasi-randomly distributed on 
the full horizontal axis (which shows for all elite cases the $K$ rank in 
logarithmic representation). Directly inspecting the numerical data 
we find that there are (for the case $w_r=0.5$) 305 nodes out of the 1000 CheiRank elite nodes 
which change polarization from blue to to red (i.e. the sign of $M$ from 
negative to blue) but otherwise their values of 
$M$ strongly fluctuate between $-1$ and $1$.

For $w_r=1$ and both elite cases we have more or less $M\approx 1$ 
for the top PageRank nodes and also strongly fluctuating values 
$-1\le M\le M$ for larger values of $K(j)$ with a preference 
for positive polarization $M>0$ in the crossover regime 
(and also very large $K$ rank values). The crossover regimes 
start roughly at $K(j)\approx 10^2$ for PageRank elite or 
$K(j)\approx 10^3$ for CheiRank elite. 

We note that there are 
nodes which were considered as 
initially blue ones and that some of them 
change their polarization 
within the network of doubled size
from blue to red. In such cases
it may be argued that their influence matrix
of (\ref{eq_sigmadef}) should also be updated from blue to red preference. 
However, this corresponds to some kind of time dependent model
which is more complicated for analytical and numerical analysis.
Therefore, we assume in this work that the memory of the original blue 
(or red) preference is preserved and such a node continues
to propagate its blue (or red) influence
with the matrix transitions 
as described in Fig.~\ref{fig1}.
The dynamical variation of influence, 
depending on the actual polarization of nodes,
will be considered in further studies.

\section{Effect of resistance in opinion formation}
\label{sec6}

Above we considered the influence matrix
described by Fig.~\ref{fig1} and (\ref{eq_sigmadef}).
In these relations it is assumed that
a node with red preference propagates 100\% red influence 
on red and blue components of other nodes. However, we can also
consider the situation in which
for the blue component there is not a 100\%
red influence but e.g. only a 80\%
influence. This means that a blue component
realizes a certain resistance to red influence
and vise verse a red component has a similar resistance
to blue influence. This is modeled by a modified
form of the transition matrices which instead of (\ref{eq_sigmadef})
take the form
\begin{equation}
\label{eq_sigmadefmodif}
\sigma_+=\left(\begin{array}{cc}
1 & 0.8 \\
0 & 0.2 \\
\end{array}\right)
\quad,\quad
\sigma_-=\left(\begin{array}{cc}
0.2 & 0 \\
0.8 & 1 \\
\end{array}\right)\ .
\end{equation}
This modification corresponds to a 20\% resistance
to influence another color. We construct the Google matrix $G_2$ in the same 
way as described in Sec.~\ref{sec2} but using the matrices 
$\sigma_\pm$ of (\ref{eq_sigmadefmodif}) to replace the unit elements of the 
adjacency matrix (of the original network). (The teleportation vector is 
the same as in Sec.~\ref{sec2}.) We call this model the modified Ising-PROF model. 
Due to the modification of the $\sigma_{\pm}$ matrices we obtain 
in (\ref{eq_PR1}) additional contributions proportional to the 
difference $P_r(j)-P_b(j)$ and the analytical argument that 
provided the relation $P_r(j)+P_b(j)=P(j)$ is no longer valid for the modified 
Ising-PROF model. 

\begin{figure}[h]
\begin{center}
\includegraphics[width=0.48\textwidth]{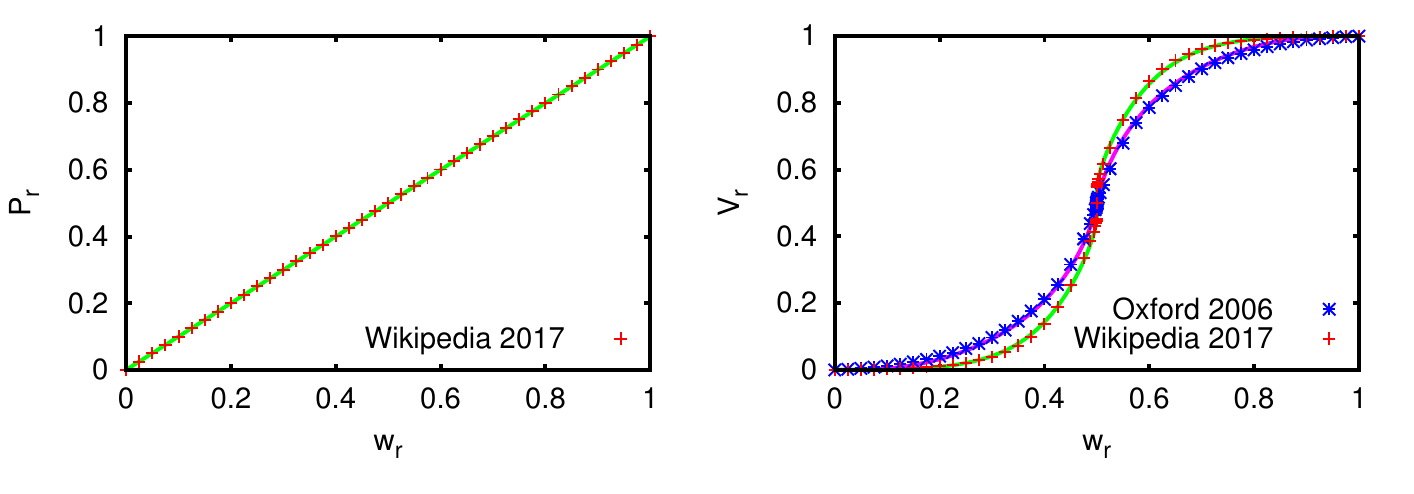}
\end{center}
\caption{\label{fig7} Same as Fig.~\ref{fig2} for the modified 
Ising-PROF model based on (\ref{eq_sigmadefmodif}); 
here in the right panel 
the full curves correspond directly to the theoretical expression 
$V_r^{\rm (th)}(w_r)$ given in (\ref{eq_Vrth}) without any rescaling.}
\end{figure}

The dependence of total red PageRank probability $P_r$
and vote $V_r$ on $w_r$ are shown in Fig.~\ref{fig7}. They are very similar 
to those of Fig.~\ref{fig2}. For $V_r$ the theoretical expression
for $V_r^{\rm (th)}(w_r)$ given in (\ref{eq_Vrth}) 
directly fits the numerical data without rescaling even though 
this theoretical expression was derived on the assumption of 
$P_r(j)+P_b(j)=P(j)$ which is no longer valid. 
We believe that this is due to statistical
fluctuations of the quantity $\tilde P(j)=P_r(j)+P_b(j)$, 
which qualitatively replaces $P(j)$ in (\ref{eq_variance1}) and 
(\ref{eq_adef}), such that the conditions to apply the central limit 
theorem are better fulfilled (for the sum of a modest number of random 
variables). Of course, this argument is not rigorous since especially 
in (\ref{eq_variance1}) the fluctuations of $\tilde P(j)$ should produce 
additional contributions which are very complicated to determine 
analytically.

\begin{figure}[h]
\begin{center}
\includegraphics[width=0.48\textwidth]{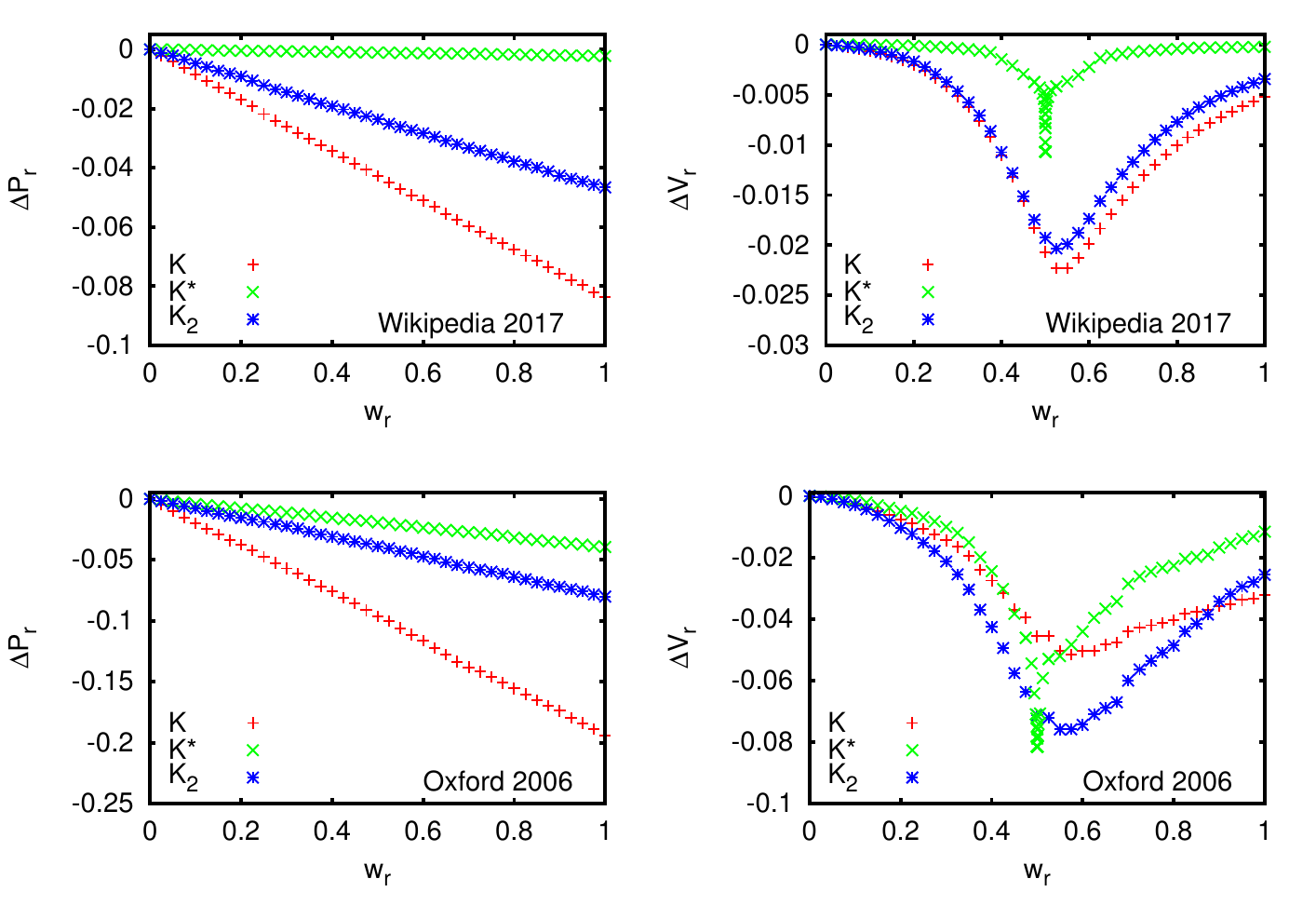}
\end{center}
\caption{\label{fig8} Same as Fig.~\ref{fig3} for the modified 
Ising network model based on (\ref{eq_sigmadefmodif}).}
\end{figure}

The elite influence for the modified Ising-PROF model
is shown in Fig.~\ref{fig8}.
We see that in this case the variation of vote induced
by elite is rather similar to the initial
Ising-PROF model. Only for Wikipedia 2017 (and 
PageRank and 2DRank elite) the maximal variation is increased
from 1.6\% for the Ising-PROF model
to 2.2\% for the modified Ising-PROF model.
On the basis of these results 
we conclude that the particular form of the influence matrices of 
(\ref{eq_sigmadef}) or (\ref{eq_sigmadefmodif}) 
does not affect the general nature of the obtained results.

\section{Discussion}
\label{sec7}

In this work we proposed the Ising-PageRank model of
opinion formation
which generates the opinion formation of
a directed social network 
using only the local information
about the neighbors of a given elector (node). 

For the homogeneous model without elite we obtain for the vote 
quantity a smooth step function as a function 
of the parameter $w_r$ and the finite effective width of the transition 
around $w_r\approx 0.5$ from $V_r=0$ (for $w_r<0.5$) 
to $V_r=1$ (for $w_r>0.5$) 
is roughly the typical value of the parameter $a_i$ given in (\ref{eq_adef})~:
\begin{equation}
\label{eq_ainv}
a_i= \frac{1}{P(i)}\sqrt{2\sum_{j\in L_i} \frac{P(j)^2}{d_j^2}}
\end{equation}
which takes an average value of about $0.3$ ($0.5$) for Wikipedia 2017 
(Oxford 2006). The right panels of Figs.~\ref{fig2} and \ref{fig7} clearly 
confirm the ratio of this effective width between the two networks and 
its overall size.

The most interesting feature of
our results  in this model is the existence of 
the strong influence of elite, which is given as a small number of top nodes
of PageRank, CheiRank or 2DRank. Even a small fraction of elite electors
produces a significant 
influence on the final vote on 
a society which is close to a 50-50
distribution of opinions between red and blue options.
Thus a small insignificant fraction
of elite nodes can push the outcome of the final vote to either a 
blue or a red majority. The variation
of vote induced by elite nodes
is expressed through the analytical relation
(\ref{eq_dVrdep}). 
We believe that the proposed Ising-PROF model can 
describe important features of opinion formation in social networks.

\section{Acknowledgments}
This work was supported in 
part by the Programme Investissements
d'Avenir ANR-11-IDEX-0002-02, 
reference ANR-10-LABX-0037-NEXT (project THETRACOM).
This work was granted access to the HPC resources of 
CALMIP (Toulouse) under the allocation 2018-P0110.

\end{document}